# Law-based and standards-oriented approach for privacy impact assessment in medical devices: a topic for lawyers, engineers and healthcare practitioners in MedTech.

## Yuri R Ladeia[1], David M Pereira[2,3]

[1] JUSGOV. Escola de Direito, Universidade do Minho. Braga. Portugal

[2] Universidade do Porto, Faculdade de Farmácia, Laboratório de Farmacognosia, Porto, Portugal.

[3] Universidade Católica Portuguesa, Escola de Direito, Porto, Portugal

**Abstract**

**Background:** The integration of the General Data Protection Regulation (GDPR) and the Medical Device Regulation (MDR) creates complexities in conducting Data Protection Impact Assessments (DPIAs) for medical devices. The adoption of non-binding standards like ISO and IEC can harmonize these processes by enhancing accountability and privacy by design.

**Methods:** This study employs a multidisciplinary literature review, focusing on GDPR and MDR intersection in medical devices that process personal health data. It evaluates key standards, including ISO/IEC 29134 and IEC 62304, to propose a unified approach for DPIAs that aligns with legal and technical frameworks.

**Results:** The analysis reveals the benefits of integrating ISO/IEC standards into DPIAs, which provide detailed guidance on implementing privacy by design, risk assessment, and mitigation strategies specific to medical devices. The proposed framework ensures that DPIAs are living documents, continuously updated to adapt to evolving data protection challenges.

**Conclusions:** A unified approach combining European Union (EU) regulations and international standards offers a robust framework for conducting DPIAs in medical devices. This integration balances security, innovation, and privacy, enhancing compliance and fostering trust in medical technologies. The study advocates for leveraging both hard law and standards to systematically address privacy and safety in the design and operation of medical devices, thereby raising the maturity of the *MedTech* ecosystem.

1. Introduction – Current state of health data protection in medical devices

From a regulatory standpoint, few, if any, fields are as densely regulated as healthcare[1]. It comprises an array of distinct sectors, from drugs to medical procedures. In all cases, highly sensitive personal data, such as health-related data is required and generated. Among the different subfields within digital healthcare and innovation, the current work focuses on medical devices.

A medical device is broadly defined as any instrument, apparatus, implement, machine, appliance, implant, reagent for *in vitro* use, software, material, or other similar or related article, intended to be used, alone or in combination, for human beings, for one or more of the specific medical purposes [2]. These purposes include the diagnosis, prevention, monitoring, treatment, or alleviation of disease; the diagnosis, monitoring, treatment, alleviation of, or compensation for an injury or disability; the investigation, replacement, modification, or support of the anatomy or a physiological process.

The processing of health-related information is foundational in healthcare [3], demanding rigorous alignment with the fundamental right to privacy, as enshrined in international treaties under Article 8 of the European Convention on Human Rights (ECHR), in effect since 1953. This right is further embodied within the European Union's (EU) data protection framework, essentially through Regulation (EU) 2016/679, the General Data Protection Regulation (GDPR), which concretises this right resulting from Article 8 of the Charter of Fundamental Rights of the European Union (CFR), proclaimed on December 2000. The GDPR, being a regulation, warrants conjugation with additional field-specific regulations. In the case of medical devices, the GDPR should be applied in tandem with Regulation (EU) 2017/745 on Medical Devices (MDR) to balance the individuals' rights and the safety of the devices. The MDR clarifies that data protection laws need to be applied when medical devices process personal data, under Recital 47 and articles 62(4)(h), 72(3), 92(4), 110(1)– (2). Therefore, if a medical device regulated by the MDR processes personal data, it also falls under the GDPR [4].

Importantly, the increased technological development of medical devices results in a steady increase of devices with digital features capable of processing health data. In this regard, according to GDPR Article 35 (1) and Recital 84, where a type of processing, in particular using new technologies, is *"likely to result in a high risk to the rights and freedoms of natural persons, the controller should be responsible for the carrying-out of a data protection impact assessment to assess, in particular, the origin, nature, particularity and severity of that risk"*, which crystalizes the desired outcomes of the process known as Privacy Impact Assessment (PIA), also officially defined as Data Protection Impact Assessment (DPIA). Although mandatory, at least for the scenarios depicted in the GDPR, the same regulation allows, under Article 35, for DPIA processes to be flexible in their formal organization and structure, provided they meet the regulatory requirements. This results in DPIAs being essentially structure and format-agnostic, provided they cover said requirements. However, such neutrality results in a lack of a systematic methodology to assess security, safety and privacy risks associated with the use of medical devices, with putative consequences to patient health [5].

To bridge these gaps, this paper proposes a unified approach to conducting DPIAs centred on medical devices with data processing capabilities, presenting a methodology to conduct DPIAs to assist law, technology, and health practitioners in assuring data security, personal data protection and privacy. The suggestion is to address the challenge by bringing together hard law such as the GDPR and MDR, with well-defined international standards – soft law – such as those emanating from the International Standard Organization (ISO) and the International Electrotechnical Commission (IEC).

Notwithstanding the importance of taking into account the prospective EU Legal Framework for the cybersecurity of medical devices and its relationships, including Data Laws (GDPR, Data Governance Act, European Health Data Space and Data Act), Medical Device Laws (MDR), AI Laws (AI Act) and Cybersecurity Laws (NIS Directive and Cybersecurity Act)[6] – for the sake of focus and space economy - this paper builds solely upon the GDPR, MDR and some selected ISOs/IECs can help build robust DPIAs in the context of medical devices.

2. **Methods**

The methodology of this paper is rooted in a multidisciplinary literature review, combining legal analysis with a review of related technical standards. The core legal frameworks examined include the GDPR and MDR (**Table 1**), with a specific focus on their intersection in regulating medical devices that process personal health data. The GDPR's Article 35, which mandates DPIAs for high-risk data processing activities, is closely analysed in the context of medical devices. In parallel, the MDR's Recitals 47 and Articles 62(4)(h), 72(3), and 110 are explored to understand the obligations imposed on medical device manufacturers regarding safety and data protection.

The review also considers international standards, particularly those issued by the ISO and IEC (**Table 1**), which provide essential technical guidance to complement the GDPR and MDR. These standards, such as ISO/IEC 29134, ISO/IEC 27701 and ISO/IEC 27002, serve as tools to implement *in concretum* data protection by design and default, as required under article 25 GDPR.

**Table 1** – Legal and technical sources used in this work – divided between non-binding (soft law) and binding rules (hard law).

| Source | Type | Description |
|---|---|---|
| **GDPR** | Hard law | General Data Protection Regulation (EU 2016/679) |
| **MDR** | Hard law | Medical Device Regulation (EU 2017/745) |
| **ISO/IEC 27001:2022** | Soft law | Information security, cybersecurity and privacy protection — Information security management systems — Requirements |
| **ISO/IEC 27002** | Soft law | Cybersecurity controls |
| **ISO/IEC 27701:2019** | Soft law | Security techniques — Extension to ISO/IEC 27001 and 27002 for privacy information management — Requirements and guidelines |
| **ISO/IEC 29134:2023** | Soft law | Information technology — Security techniques — Guidelines for privacy impact assessment |
| **ISO/IEC 29151:2017** | Soft law | Information technology — Security techniques — Code of practice for personally identifiable information protection |
| **IEC 62304:2006** | Soft law | Medical device software — Software life cycle processes |

## 3. Results and discussion

*3.1. An overview of medical devices architecture and their regulation*

As we mentioned earlier, a medical device is broadly defined as any instrument, apparatus, implement, machine, appliance, implant, reagent for *in vitro* use, software, material, or other similar or related article, intended to be used, alone or in combination, for human beings, for one or more of the specific medical purposes [1]. All these purposes can be found in detail in the relevant legislation, namely the MDR (Article 2) and the In Vitro Diagnostic Medical Devices Regulation (IVDR, EU Regulation 2017/746). An important distinction is that, while medicines and drugs attain their preventive, therapeutic or diagnostic by pharmacological means, medical devices do not [7], although there are some mixed-type medical devices that release drugs or medicines [8], which are outside the scope of this work.

Finally, it should be mentioned that the thousands of medical devices in existence are, *as per* the MDR, classified according to their risk profile, varying from class I to class III. This medical device categorization is based on the risk they pose to patients and users. The classification determines the level of regulatory scrutiny they must undergo before entering the market.

Class I MD (Low Risk) include low-risk devices that do not come into contact with vital organs or systems. They are typically non-invasive and may include items used temporarily or for simple monitoring, such as bandages, examination gloves or band-aids. Class I devices are often exempt from premarket notifications, owing to their limited technology integration and minimal risk to patients. Such criteria take into account factors such as the duration of use, invasiveness, and the criticality of the device's function.

Class II (further divided in class IIa and IIb, moderate to high risk) comprise devices that are generally more complex than class I devices and may be invasive or involve prolonged use. Here, we find devices such as surgical needles, dental fillings, and hearing aids (class IIa), and also infusion pumps, ventilators, and certain diagnostic imaging equipment (class IIb).

Finally, class III MD (high risk) are often life-supporting or life-sustaining and present a high potential risk to human life in case of failure. They include MD such as heart valves, pacemakers, and implantable defibrillators. Relevantly, a similar system is used in the USA, where MD are also classified into three classes (I, II, III) based on similar risk factors at the federal level (Food, Drug, and Cosmetic Act, US Code, Title 21, Chapter 9, Subchapter V, sections 360c to 360k)

From a legal point of view, this classification is very important as the classification of medical devices not only influences the regulatory pathways for approval to the EU market but also affects post-market obligations, including monitoring, reporting of adverse events, and ensuring ongoing compliance with data protection. Additionally, DPIAs are not required in the case of medical devices that are devoid of data processing features.

Considering the classification of medical devices according to their level of risk, if their outputs depend on data processing, the quality and security of this data is decisive for the patient's safety. Indeed, data protection and product safety go hand in hand and are interdependent.

*3.2. Data protection challenges associated with the use of medical devices*

Notwithstanding the current use of thousands of medical devices, most of which still devoid of data processing capabilities, the field witnessed a growing trend of increasingly sophisticated products [9], often integrating advanced technology to provide critical health insights and therapeutic benefits. However, such advancements bring challenges, particularly regarding data protection. Indeed, these devices process sensitive personal data which includes main operations such as collection, storage, transfer or erasure, as per Article 4 (2) GDPR.

Manufacturers must engage and implement measures to ensure data protection falls into several categories, such as technical and organizational measures. Likewise, data protection is to be assured both by default and also by design (GDPR, Article 25), frequently requiring technical and organizational measures. Such measures are typically including in the DPIA, which not only maps the data processing activities involved in the use of the device, but also the identified risks and mitigation controls.

Failure to conduct a DPIA when required – before data processing – or to adequately address the risks identified, can result in significant fundamental rights violations, commercial damages with civil and criminal liability compensation, fines and penalties under the GDPR and national member states related laws. Under GDPR, these can include fines of up to €10 million or 2% of the global annual turnover, whichever is higher (GDPR, Article 83[4][a]). Indeed, there have been some decisions from European Courts regarding violation of personal data processing [10], including health data [11] .

*3.3.Proposed risk assessment framework for DPIAs*

As we mentioned earlier, there is no mandated structure or template for a DPIA. However, the former Article 29 Working Party on Data Protection – now the European Data Protection Board (EDPB), has produced guidelines, still valid, that stipulate the key information necessary for a DPIA, as well as its corresponding minimum requirements [12], as depicted in **Table 2**. Although the open structure of a DPIA allows it to be adjusted to particular characteristics and needs of distinct technical fields, in the specific case of medical devices, which are subjected to numerous and tight regulations, such liberty often results in markedly different DPIA structures across the field. To this end, ISO/IEC standards can help organizations achieve greater consistency and reliability in their privacy risk assessments. This approach not only aligns with global best practices for privacy management but also facilitates compliance with various regulatory requirements across jurisdictions. Leveraging these standards ensures that privacy considerations are systematically integrated into the design and operation of medical devices, thereby enhancing the protection of personal data and reducing the risk of regulatory breaches and associated liabilities.

Another advantage of using such standards is that they can be updated and keep up with technological developments more easily than the legislative process can normally produce normative content. For example, the ISO/IEC 29134:2017 standard was updated to ISO/IEC 29134:2023 [13] just six years after it came into force, due to the rapid growth in the field of security techniques it addresses.

**Table 2** – Major steps to be observed in a DPIA, corresponding scope and objectives, and suggested standards.

| STEPS | SCOPE/OBJECTIVES | STANDARDS |
|---|---|---|
| **Processing Description**<br><br>GDPR, Recital 90<br>GDPR, Article 35(7)(a)<br>GDPR, Article 35 (8) | Nature, scope, context and purposes of the processing considered;<br>Personal data, recipients, and storage duration recorded;<br>Functional description of the processing operation provided;<br>Assets (hardware, software, networks, people, paper, etc.) identified;<br>Compliance with approved codes of conduct | ISO/IEC 29134:2023 [13]<br>IEC 62304:2006 [14] |
| **Necessity & Proportionality**<br><br>GDPR, Article 35(7)(b/d)<br>GDPR, Recital 90<br>GDPR, Article 5(1)(b/c/e)<br>GDPR, Article 6 | Measures to comply with the Regulation determined;<br>Specified, explicit, and legitimate purpose(s);<br>The lawfulness of processing;<br>Adequate, relevant, and limited data<br>Limited storage duration | ISO/IEC 27701:2019 [15]<br>ISO/IEC 29151:2017 [16] |
| **Data Subject Rights**<br><br>GDPR, Articles 12-17<br>GDPR, Articles 19-20 | Information provided to data subjects;<br>Right of access and data portability;<br>Right to rectification and erasure;<br>Right to object and restrict processing; | ISO/IEC 29134:2023 [13]<br>ISO/IEC 29151:2017 [16]<br>ISO/IEC 27002:2022 [17]<br>ISO/IEC 27701:2019 [15] |
| **Processor Relations** | Relationships with processors;<br>Safeguards for international transfers; | ISO/IEC 29134:2023 [13] |

| | Article 28 | | |
| --- | --- | --- | --- |
| | Article 44-50 | | |
| **Risk Management**<br><br>GDPR, Article 25<br>GDPR, Article 32<br>GDPR, Article 35(7)(c/d)<br>GDPR, Recital 84<br>GDPR, Recital 90 | Risks to rights and freedoms of data subjects managed;<br>Origin, nature, particularity, and severity of risks assessed;<br>Risk sources considered;<br>Potential impacts identified<br>Threats leading to risks identified<br>Likelihood and severity of risks estimated;<br>Measures to treat risks determined; | ISO/IEC 27001:2013 [18]<br>ISO/IEC 29151:2017 [16]<br>IEC 62304:2006 [14]<br>ISO/IEC 29134:2023 [13] | |
| **Interested parties**<br><br>GDPR, Article 35(2)<br>GDPR, Article 35(9) | Advice from the Data Protection Officer (DPO);<br>Views of data subjects or their representatives sought where appropriate; | ISO/IEC 29134:2023 [13] | |
| **Document and approve the DPIA**<br><br>GDPR, Articles 35/36 | Document the DPIA process, including all identified risks and the measures implemented to address them.<br><br>The DPIA might need to be subject of Data Protection Officer (DPO) advise and, in residual high-risk cases, submitted to the prior consultation with Data Protection Authority (DPA). | ISO/IEC 29134:2023 [13]<br>ISO/IEC 29151:2017 [16] | |
| **Monitor and Review**<br>GDPR, Article 35(11) | DPIAs should be considered living documents, subject to review and updating as processing activities or risks change. Regular monitoring ensures that the measures remain effective. | ISO/IEC 29134:2023 | |

For example, the initial step of a DPIA, in which processing operations are described, is pivoted in GDPR Article 35 as a starting point. Such statute, however, can benefit from the concomitant use of ISO/IEC 29134:2023 [13], which outlines the necessary components for describing data processing activities. This includes detailing the nature, scope, context, and purposes of processing. The standard emphasizes the importance of clearly identifying and documenting the data types being processed, the stakeholders involved (e.g., data controllers, processors, and data subjects), and the flow of data within the system. In the specific case of medical devices that include software components, IEC 62304:2006 [14] provides a framework for the software life cycle processes specific to medical device software. In the context of describing the processing operation, this technical document guides the definition of the software system, including its intended purpose, functions, and how it interacts with other systems and data flows. This is critical for accurately describing the data processing activities of the software component within a medical device.

Another step, assessing necessity and proportionality in data processing activities, is largely based on the dispositions of GDPR articles 5, 6 and 35. They can, however, benefit from the concomitant use of ISO/IEC 27701:2019 [15], which guides on implementing data minimization and purpose limitation, pivotal resources to assessing necessity. The standard encourages organizations to only collect and process data that is directly necessary for the specified purposes, helping align with the GDPR's requirement to evaluate necessity. By implementing these controls, organizations can ensure that data processing is not excessive and is strictly limited to what is necessary to achieve the intended purpose. Further contributions can be drawn from ISO 29151:2017 [16], which provides guidelines for documenting and justifying the need for processing activities.

For step 3 of a DPIA, which involves identifying and assessing risks, GDPR Article 35 provides the foundation by mandating that organizations evaluate the potential impacts of data processing on the rights and freedoms of individuals. This step is crucial for understanding the privacy risks associated with processing activities and is further enhanced by the guidance offered in ISO/IEC 29134:2023 [13]. The standard elaborates on methods to identify privacy risks, emphasizing a systematic approach to evaluating both the likelihood and severity of risks. It includes recommendations for conducting risk assessments that consider factors such as unauthorized access, data breaches, and potential discrimination, thus ensuring a comprehensive evaluation of privacy impacts. Moreover, when dealing with medical devices that incorporate software, ISO/IEC 29151:2017 [16] is particularly relevant as it provides a detailed set of controls for protecting personally identifiable information (PII). This includes specific measures for assessing the risks to PII, focusing on how data is collected, stored, and transmitted. The standard emphasizes not only the identification of risks but also the need to document these risks clearly, including the contexts in which they arise and their potential impacts on individuals. This detailed documentation is essential for substantiating the risk assessments required under GDPR and for aligning with industry best practices in data protection and cybersecurity.

Once risks have been identified and mapped, step 4 of the DPIA focus the identification of measures to mitigate said risks. GDPR Articles 25 and 32 provide a legal framework for implementing appropriate technical and organizational measures to ensure data security and minimize risks. This includes adopting strategies such as encryption, pseudonymization, access controls, and regular audits. However, leveraging ISO/IEC 29134:2023 [13] alongside these legal requirements can further refine this process by offering specific guidelines on risk mitigation strategies tailored to the context of data protection impact assessments. For medical devices that involve software components, IEC 62304:2006 [14]and ISO/IEC 29151:2017 [16] provide additional critical insights. IEC 62304:2006 outlines software development processes that include safety classification and risk management, emphasizing the integration of risk control measures directly into the software life cycle. This ensures that software-related risks, including those associated with data processing, are mitigated from the design phase onwards. ISO/IEC 29151:2017 complements this by providing specific controls focused on protecting PII, such as implementing robust access control mechanisms, ensuring data accuracy, and maintaining data integrity.

The subsequent step of the DPIA process, step 5, involves the documentation and approval of the DPIA, ensuring that all identified risks and the corresponding mitigation measures are thoroughly recorded and reviewed. GDPR Articles 35 and 36 outline the requirements for documenting the DPIA process and, in certain high-risk cases, submitting the DPIA to the relevant Data Protection Authority (DPA) for consultation. This documentation serves as evidence of compliance and helps to demonstrate that the organization has taken appropriate steps to address data protection risks. ISO/IEC 29134:2023 [13] provides a structured approach for documenting DPIAs, including recommendations for the format and content of the assessment report. The standard emphasizes the importance of clear and comprehensive documentation of each stage of the DPIA, including descriptions of the processing activities, identified risks, mitigation measures, and the decision-making process. This ensures that the DPIA is not only a tool for risk management but also a transparent record that can be reviewed and audited. For medical devices with software components, IEC 62304:2006 [14] and ISO/IEC 29151:2017 [16] further reinforce the importance of thorough documentation. IEC 62304:2006 specifies the need for maintaining detailed records throughout the software development life cycle, including risk management activities and decisions regarding software safety. ISO/IEC 29151:2017 supports these requirements by providing guidelines on documenting privacy controls and their effectiveness, ensuring that all measures to protect PII are clearly articulated and justified. Additionally, ISO/IEC 27701:2019 [15] extends the documentation requirements to include privacy-specific considerations within the context of a Privacy Information Management System (PIMS). It underscores the need for clear approval processes and, where necessary, consultation with internal and external stakeholders, including Data Protection Officers (DPOs) and data protection authorities (DPA).

Step 6 of the DPIA process involves monitoring and reviewing the DPIA on an ongoing basis to ensure that the data protection measures remain effective and relevant as processing activities or associated risks evolve. GDPR Article 35(11) emphasizes that

DPIAs should be dynamic documents that are regularly reviewed and updated, especially when there are changes in the nature, scope, context, or purposes of the processing. This ensures that the DPIA continues to address current risks and remains aligned with the principles of data protection by design and by default. ISO/IEC 29134:2023 [13] supports this ongoing review by recommending regular assessments of the DPIA's effectiveness and the adequacy of implemented controls. It guides establishing review schedules and criteria, such as changes in technology, new types of data processing, or the emergence of new threats. This ensures that DPIAs are not static but evolve in response to the changing risk landscape. For medical devices with software components, IEC 62304:2006 [14] and ISO/IEC 29151:2017 [16] further highlight the importance of continuous monitoring and updating of risk management processes. IEC 62304:2006 mandates a continuous software maintenance plan that includes procedures for ongoing risk management throughout the software's life cycle. This includes monitoring for new vulnerabilities, updating software as needed, and reassessing the associated risks, which is crucial for devices that handle sensitive health data. Additionally, ISO/IEC 27701:2019 [15] extends the concept of ongoing monitoring and review within the framework of a PIMS. It recommends regular audits and continuous improvement cycles to ensure that privacy controls remain effective and compliant with evolving legal and regulatory requirements. This approach is particularly important for medical devices, where the rapid pace of technological innovation can introduce new privacy and security challenges.

Ubiquitously and *omnipresently* throughout the process, DPIA involves ensuring data protection by design and by default, a principle firmly rooted in GDPR Article 25. This article requires that data protection measures are embedded into the design of processing activities from the outset, ensuring that personal data is protected throughout the data lifecycle. ISO/IEC 29134:2023 supports this requirement by emphasizing the need to incorporate privacy controls early in the development process, specifically recommending that privacy considerations be integrated into the architecture and design phases of the data processing system. For medical devices with software components, ISO/IEC 27701:2019 and ISO/IEC 29151:2017 provide additional frameworks that are crucial for embedding privacy by design and by default. ISO/IEC 27701 extends the ISO/IEC 27001 and 27002 standards to include privacy-specific controls, guiding organizations on how to implement and maintain a PIMS that aligns with the principles of data protection by design. This involves the adoption of privacy-enhancing technologies and practices, such as pseudonymization and anonymization, to minimize data exposure and mitigate risks. Additionally, ISO/IEC 29151:2017 offers detailed controls for ensuring that data protection measures are consistently applied across all processing activities, including the use of default settings that favour privacy. These standard stresses the importance of designing systems that limit data collection, processing, and retention to what is necessary for the specified purposes, thereby operationalizing the GDPR's requirements for data protection by default. For medical device software, integrating these standards helps ensure that privacy is not an afterthought but a foundational element of the system's design and operation, thereby enhancing compliance and safeguarding personal data throughout the device's lifecycle.

By integrating these standards, organizations can ensure that their DPIAs are living documents that are continuously updated and refined, maintaining a proactive approach to data protection and ensuring compliance with both GDPR and best practices in the industry.

## 4. Conclusion

This paper explores the intricate interplay between hard law, such as the GDPR and MDR, and soft law standards, particularly those from ISO and IEC, in the context of conducting DPIA for digital medical devices. The analysis underscores that while GDPR provides a robust legal framework for data protection, its broad and sector-agnostic nature necessitates the support of specific standards that can address the unique challenges posed by medical devices, especially those with software components.

It is argued that standards such as ISO/IEC 29134:2023, ISO/IEC 27701:2019, and ISO/IEC 29151:2017 within the DPIA process to enhances the practical implementation of privacy by design and by default. Therefore, it is argued that these standards offer detailed guidance that complements the hard law under GDPR's requirements, providing a structured approach to risk assessment, risk mitigation, and the integration of privacy controls throughout the lifecycle of medical devices. Moreover, standards such as IEC 62304:2006 play a critical role in defining the software development processes and risk management specific to medical device software, ensuring that safety and privacy considerations are embedded from the outset.

Ultimately, this paper advocates for a unified, law-based and standards-oriented approach to conducting DPIAs for medical devices, highlighting the need for continuous adaptation and alignment of legal and technical measures. By leveraging both hard law and sector-specific standards, stakeholders can ensure that privacy considerations are systematically integrated into the design and operation of medical devices, thereby safeguarding personal data and fostering trust in digital health innovations. This integrated approach not only enhances compliance but also positions organizations to better navigate the complex regulatory landscape and address the emerging challenges of data protection in healthcare.

The convergence of hard law and soft law is not merely complementary but essential for achieving effective data protection in the rapidly evolving landscape of digital health technologies. As the paper demonstrates, the symbiotic relationship between these legal and technical frameworks enables a more responsive and objective approach to data protection that is capable of keeping pace with technological advancements. This is particularly crucial in the medical devices sector, where the balance between innovation, security and privacy must be expertly maintained.

In conclusion, the complex interplay between data protection, product safety, and privacy in the realm of medical devices, demands a harmonized regulatory approach. By integrating the GDPR, MDR, and ISO/IEC standards, a comprehensive framework for DPIAs can be established, ensuring that privacy by design and default principles are effectively implemented. This multidisciplinary strategy not only mitigates risks but also bridges the gap between legal requirements and the practical realities faced by manufacturers and regulators. Ultimately, this convergence fosters both innovation and compliance, safeguarding patient data and promoting trust in medical device technologies.

**Author's contribution:** Conceptualization: YR and DMP; Research: YR and DMP; Drafting: YR and DMP; Final approval: YR and DMP

**Acknowledgements**: The work was supported through projects UIDB/50006/2020, UIDP/50006/2020 and research grant 2023. 04954.BD funded by FCT (Fundação para a Ciência e Tecnologia, Portugal) /MCTES (Ministério da Ciência, Tecnologia e Ensino Superior) through national funds.

**Conflicts of Interest**: The authors declare no conflicts of interest